\documentclass[runningheads]{llncs}

\usepackage[utf8]{inputenc} 
\usepackage[T1]{fontenc}    
\usepackage{hyperref}       
\usepackage{url}            
\usepackage{booktabs}       
\usepackage{amsfonts}       
\usepackage{nicefrac}       
\usepackage{microtype}      
\usepackage{multirow}
\usepackage{amsmath}
\usepackage[caption=false]{subfig}

\usepackage{graphicx}

\usepackage[dvipsnames]{xcolor}

\usepackage{algorithm,algcompatible,amsmath}
\algnewcommand\INPUT{\item[\textbf{Input:}]}%
\algnewcommand\OUTPUT{\item[\textbf{Output:}]}%

\begin{document}
\title{Never `Drop the Ball' in the Operating Room: \ \  An efficient 
hand-based VR HMD controller interpolation algorithm, for 
collaborative, networked virtual environments} 

\titlerunning{Never `Drop the Ball' in the Operating Room}

\author{%
Manos Kamarianakis\inst{1,2}\orcidID{0000-0001-6577-0354}
\and \\ 
Nick Lydatakis\inst{1,2}\orcidID{0000-0001-8159-9956}
\and \\ 
George Papagiannakis\inst{1,2}\orcidID{0000-0002-2977-9850}}
\authorrunning{M. Kamarianakis, N. Lydatakis, G. Papagiannakis}
\institute{University of Crete, Greece \and ORamaVR, \url{http://www.oramavr.com} \email{\{manos,nick,george.papagiannakis\}@oramavr.com}}


\maketitle

\begin{figure}[htbp]
  \centering
  \subfloat[]{\includegraphics[width=0.32\textwidth]{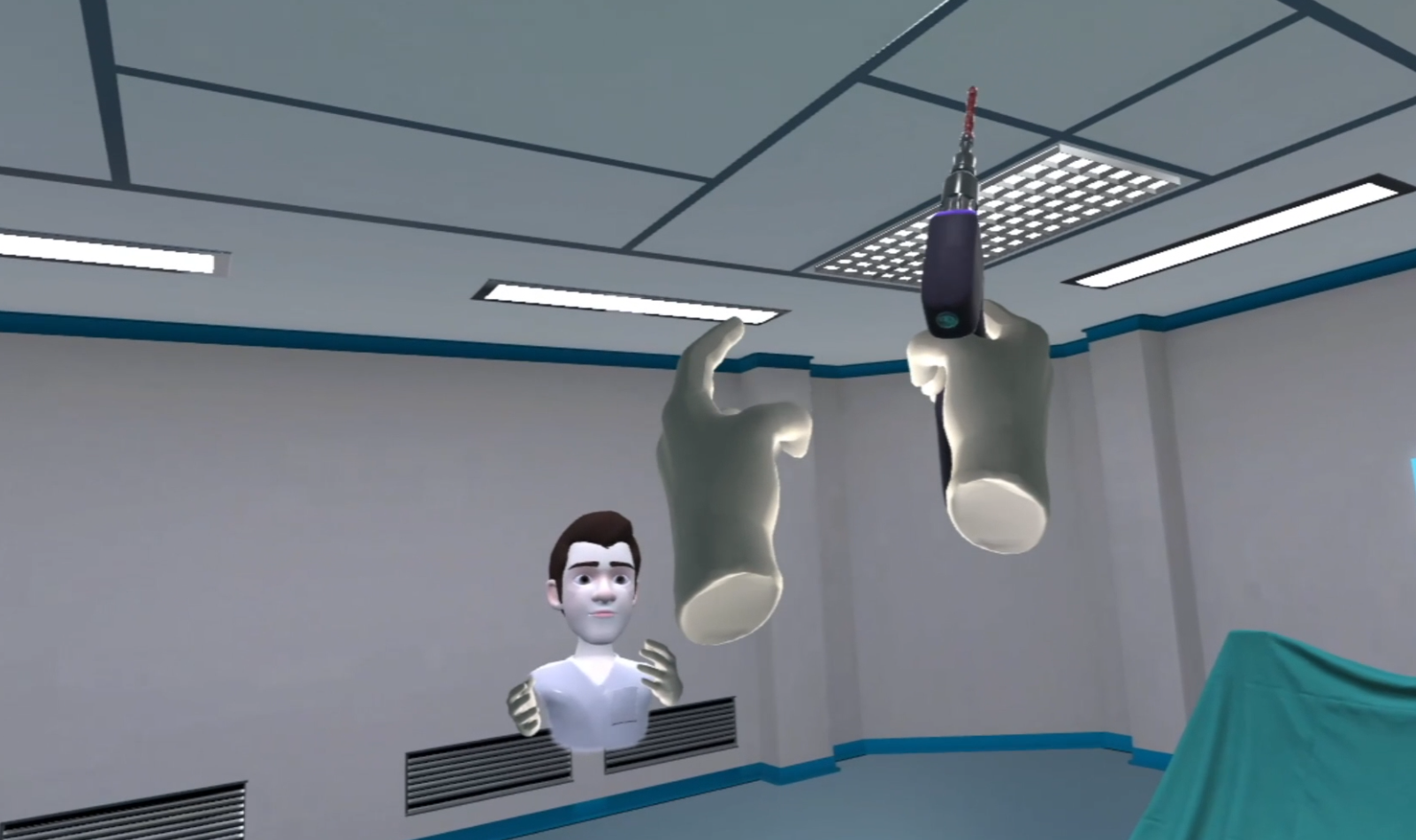}}%
  \hspace{3pt}%
  \subfloat[]{\includegraphics[width=0.32\textwidth]{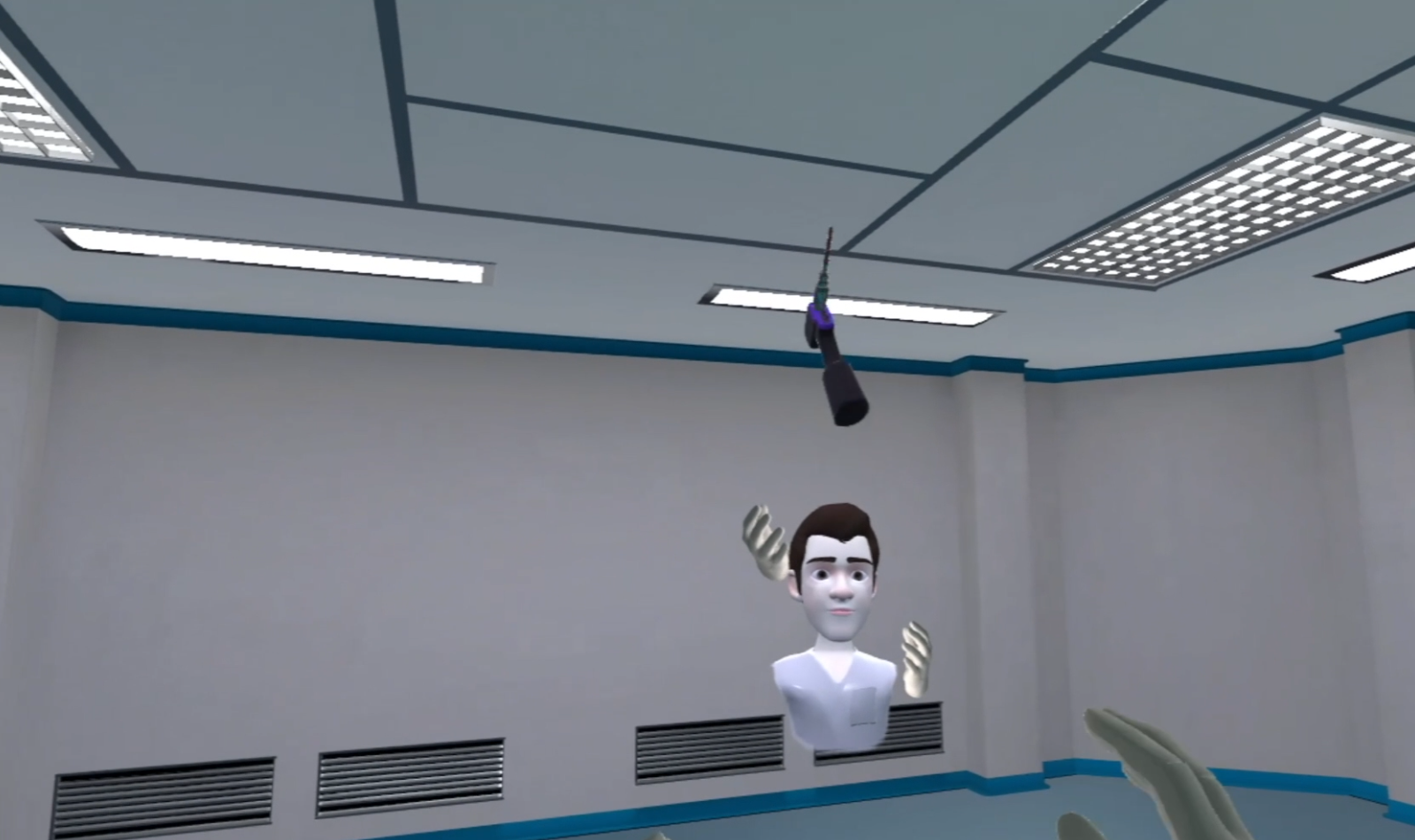}}%
  \hspace{3pt}%
  \subfloat[]{\includegraphics[width=0.324\textwidth]{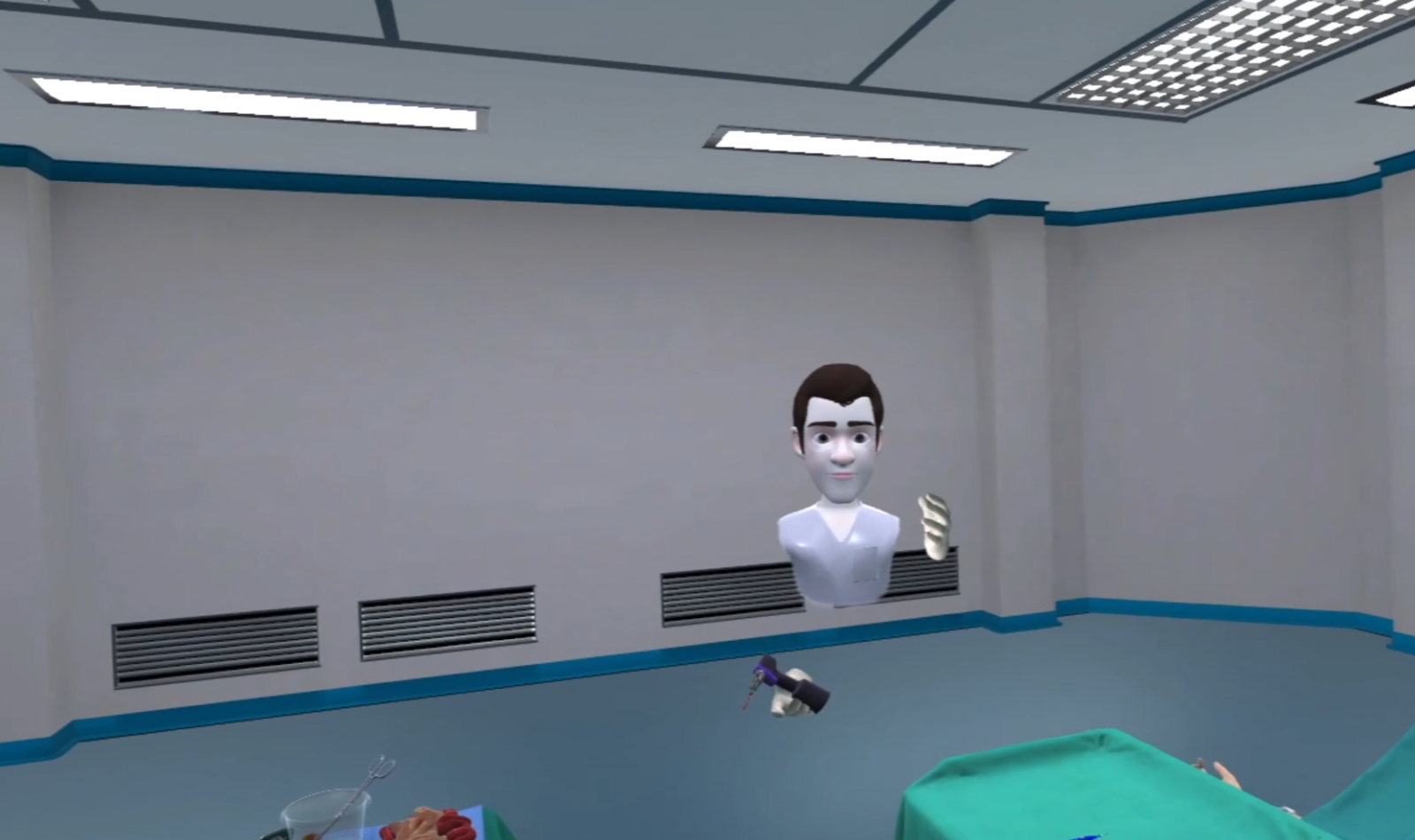}}%
  \caption{Catching a tool in a VR collaborative scenario. 
  (a) A users throw a tool (in our case a medical drill) at another. 
  (b) The objects keyframes, sent by the user  
  that threw it, are interpolated using multivector LERP 
  (see Section~\ref{sub:proposed_method_based_on_multivectors}) 
  on the receiver's VR engine. (c) The receiver manages to 
  catch the tool, as a result of the effective frame generation 
  that is visualized in his/her HMD. This example is just to 
  illustrate extreme hand-based interpolation in collaborative, 
  networked virtual environments and is provided for illustration 
  only. DO NOT TRY THIS AT HOME.}
  \label{fig:catching}
\end{figure}

\begin{abstract}
In this work, we propose two algorithms that can be applied 
in the context of a networked virtual environment
to efficiently handle the interpolation of displacement data 
for hand-based VR HMDs. Our algorithms, based on the use of 
dual-quaternions and multivectors respectively, impact the 
network consumption rate and are highly effective in 
scenarios involving multiple users. We illustrate convincing 
results in a modern game engine and a medical VR 
collaborative training scenario. 

\keywords{Interpolation  \and Keyframe Generation \and Geometric Algebra.}
\end{abstract}

\section{Introduction} 
\label{sec:introduction}

Collaborative, shared virtual environments (CVEs)
are among the most researched and developed 
areas of the last decades 
\cite{churchill1998collaborative,molet1999anyone,papagiannakis2008survey,ruan2021networked}. The growing need of remote 
networked communication, 
further accelerated by the ongoing pandemic, resulted in great 
leaps in technological advancements. 
Head-mounted displays (HMDs) are now capable of supporting intensive 
resource-demanding Virtual Reality (VR) applications.
To further facilitate this support, powerful 
algorithms are being developed and optimized by VR specialists. 

Part of this research revolves around the efficient relay
of synchronized, networked information from the HMD to the VR engine 
that is responsible for 
the rendering of the scene \cite{vilmi2020real}. 
This information typically involves user interactions 
through the HMD controllers such as \emph{displacement data} 
(e.g., translation and rotation of the controller) 
within specific time intervals and button-press
events.

Specifically, when the user moves the hand-based
controllers of his HMD, the hardware 
initially detects the movement type and logs it, in various 
time intervals based on the user's or developer's preferences. 
This logged movement, that is either a translation and/or a rotation, 
is constantly transcoded into a suitable format and relayed 
to the VR application and rendered as a corresponding 
action, e.g., hand movement, object transformation or some action. 
The controller's data format to be 
transmitted to the rendering engine affects the overall 
performance and  quality of experience (QoE) and poses challenges 
that must be addressed.
These challenges involve keeping the latency between the movement 
of the controller and its respective visualization in the 
HMD below a certain threshold that will not break the user's  
immersiveness. Furthermore, the information must be 
relayed efficiently such that a continuous movement of the controller
results in a smooth jitter-less outcome in the VR environment. 
Such challenges heavily depend on the implementation details 
regarding the communication channel that handles the 
way that position and rotation of the controller is relayed, as well 
as the choice of a suitable interpolation technique. After all, 
the displacement data are transmitted at discrete time
intervals, approximately 20-40 times per second. To maintain a high 
frame-per-second scenery on the VR, multiple in-between frames 
must be created on-the-fly by the appropriate tweening algorithm. An efficient 
algorithm will allow the generation of natural flow frames 
while requiring less intermediate keyframes. 
Such algorithms will help reduce a)
bandwidth usage between the HMD and the rendering engine and
b) CPU-strain, 
resulting in lower energy consumption as well as lower latency issues 
in bandwidth-restricted networks. Moreover, HMDs with 
controllers of limited frequency will still be able to deliver 
the same results as more expensive HMDs. 

The current state-of-the-art methods regarding the format used to 
transmit the displacement data mainly involves the use of 3D vectors for 
translation and quaternions for rotation data. These representation 
forms are dominant due to the fact that they involve very few bytes 
to be represented (3 and 4 respectively) and the fact that they 
support fast and efficient interpolations. Specifically, 3D vectors
are usually linearly interpolated where as the SLERP method is 
usually used for quaternion blending. In some engines, such as 
Unity3D, rotations are sometimes provided in terms of Euler angles, but 
for interpolation needs, they are internally transformed to 
their quaternion equivalents. 

\begin{figure}[htbp]
  \centering
  \includegraphics[width=0.31\textwidth]{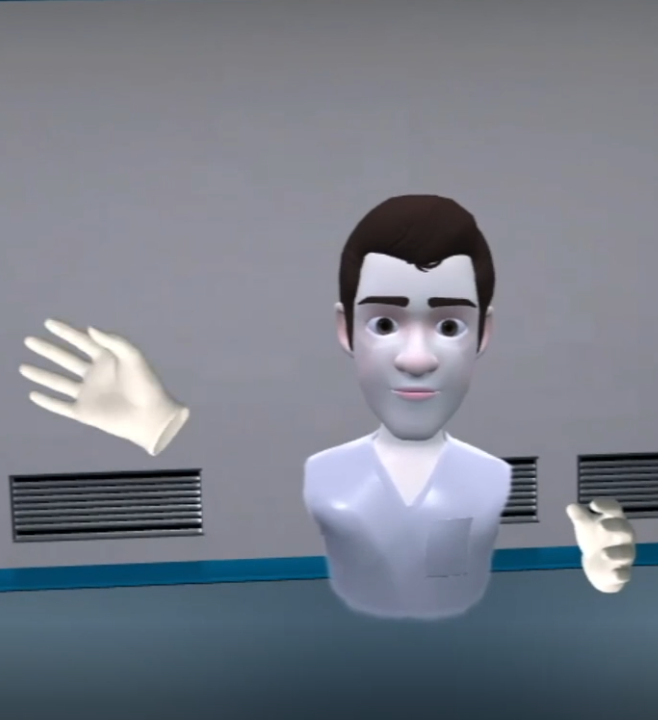} 
  \includegraphics[width=0.33\textwidth]{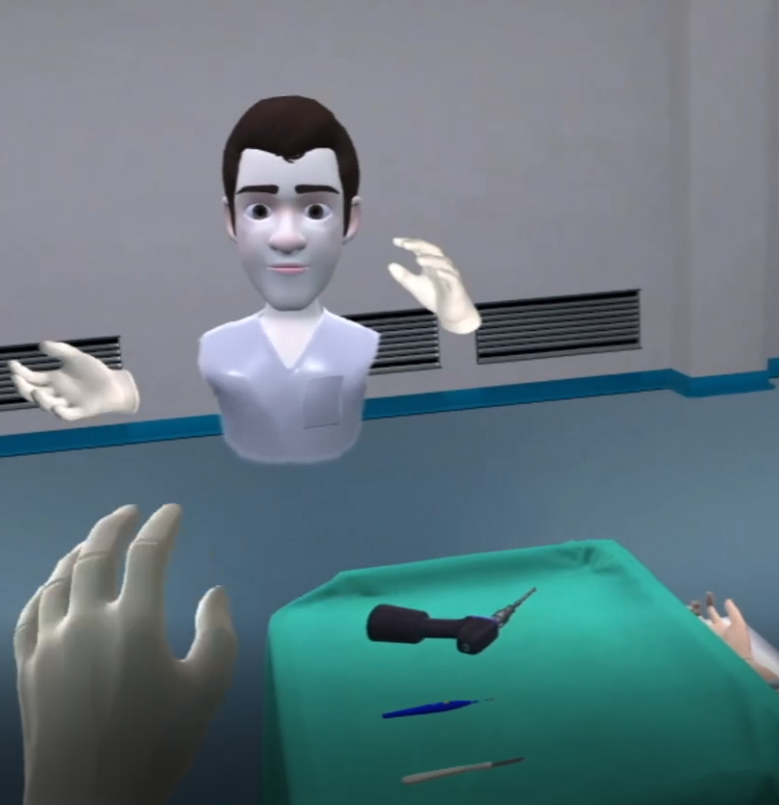}
  \includegraphics[width=0.31\textwidth]{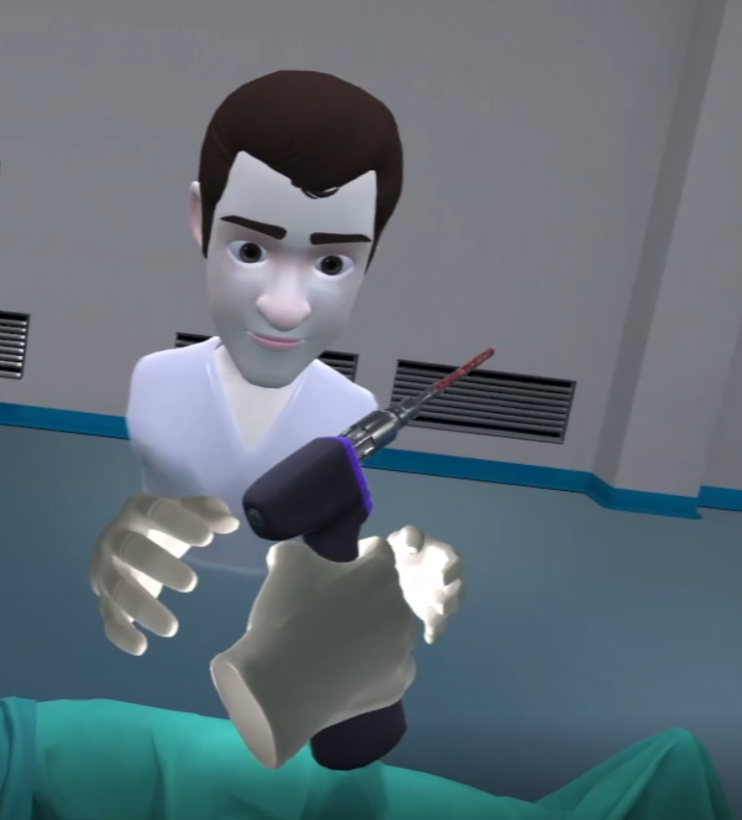}

  \caption{Images taken from a modern VR training application 
  that incorporates our proposed interpolation methods 
  for all rigid object transformations as well as hand 
  and avatar movements. }
  \label{fig:flexing}
\end{figure}

\textbf{Our Contribution.} In this work, we propose 
the use of geometric algebra (GA) as a mean to encapsulate the 
positional and rotational data of the hand-based VR HMD
controllers and to generate 
the intermediate frames in the rendering engine.
Our idea aims to take advantage of the fact that basic geometric 
entities used in VR, such as points, planes, lines, translations, 
rotations and dilations (uniform scalings), can be uniformly 
represented as \emph{multivectors}, i.e., elements of a suitable 
geometric algebra such as 3D Projective (3D PGA) or 
3D Conformal Geometric Algebra (3D CGA). 
Algebras such as 3D PGA and 3D CGA are showing rapid adaptation to VR
implementations due to their ability to represent the commonly used
vectors, quaternions and dual-quaternions natively as multivectors. 
In fact, quaternions and dual-quaternions are contained as 
a sub-algebra in both these algebras\cite{DietmarFoundations}. 
Therefore, they incorporate 
all benefits of quaternions and dual-quaternions representation such 
as artifact minimization in interpolated frames\cite{Kavan2008}. 
 Furthermore, 
geometric algebras enable powerful geometric predicates and modules
within an all-in-one framework \cite{kamarianakis2021all}, providing, 
if used with caution, performance which is on par with the current state-of-the-art frameworks\cite{Papaefthymiou:2016dx}. 
We illustrate convincing results in a modern game engine and a 
medical VR collaborative training scenario (see 
Figure~\ref{fig:flexing} and the video accompanying this video, 
found in \url{https://bit.ly/3gqeera}).

\section{State of the Art} 
\label{sec:state_of_the_art}


The current state-of-the-art for representing the controllers 
displacement are 3D vectors for the positional data and quaternions 
for the rotational data. Regarding the position, the controllers 
log their current position $v=(v_x,v_x,v_z)$ at each time step with 
respect to a point they consider as origin. 
Their rotation is stored as a 
\emph{unit} quaternion $q=(q_x,q_y,q_z,q_w)=q_xi+q_yz+q_zk+q_w$, i.e., it 
holds that $q_x^2+q_y^2+q_z^2+q_w^2=1$. The use of unit quaternions 
revolutionized graphics as it 
provided a convenient, minimal way to represent rotations, 
while avoiding known problems (e.g., gimbal lock) 
of other representation forms such as Euler angles \cite{Kavan2008}. 
The ways to transmute between unit quaternions and 
other forms representing the 
same rotation, such as rotation matrices and Euler angles, 
are summarized in \cite{diebel2006representing}. 

The interpolation of the 3D vectors containing the positional data 
is done linearly, i.e., given $v$ and $w$ vectors we may generate 
the intermediate vectors $(1-a)v+aw$, for as many $a\in [0,1]$ as 
needed. Given the unit quaternions $q$ and $r$ the intermediate 
quaternions are evaluated using the SLERP blending, i.e., we evaluate
$q(q^{-1}r)^a$, for  as many $a\in [0,1]$ as 
needed, like before. If these intermediate quaternions are applied 
to a point $p$, the image of $p$, as $a$ goes from $0$ to $1$, 
has a uniform angular velocity around a fixed rotation axis, which 
results to a smooth rotation of objects and animated models.

\section{Room for Improvements} 
\label{sec:room_for_improvements}

  The current state for representing and interpolating 
  positional and rotational data is based on the use of 3D vectors 
  and quaternions as the main VR rendering engines, Unity3D and Unreal 
  Engine, have the respective frameworks already built in. 
  Graphics courses worldwide mention quaternions as the next evolution 
  step of Euler angles; a step that simplified things and 
  amended interpolation problems without adding too much overhead 
  in the process. Despite it being widespread, the combined use 
  of vectors and quaternions does not come without limitations. 

  A drawback that often arises lies on the fact that the 
  simultaneous linear interpolation of the vectors with the SLERP 
  interpolation of the quaternions applied to rigid objects does not 
  always yield smooth, natural looking results in VR. This is empirically 
  observed on various objects, depending on the movement the user 
  \emph{expects} to see when moving the controllers. Such  
  \emph{artifacts} usually require the developer's intervention 
  to be amended, usually by demanding more intermediate 
  displacements from the controller to be sent, i.e., by 
  introducing more non-interpolated keyframes. This results mainly 
  in the increase of bandwidth required as more information 
  much be sent back and forth between the rendering engine and the 
  input device, causing a hindrance in the networking layer. 
  Multiplayer VR applications, that heavily rely on the  
  input of multiple users on the same rendering engine for multiple 
  objects, are influenced even more, when such a need arises. 
  Furthermore, the problem is intensified 
  if the rendering application resides on a cloud or edge node; 
  such scenarios are becoming increasingly more common as they 
  are accelerated by the advancements of 5G networks and the 
  relative functionalities they provide. 

\section{Proposing New Approaches} 
\label{sec:proposing_new_approaches}

  \subsection{Proposed Method Based on Dual Quaternions} 
  \label{sub:proposed_method_based_on_dual_quaternions}

  In the past few years, graphics specialists have shown that 
  dual quaternions can be a viable alternative and improvement 
  over quaternions, as they allow us to unify the translation and 
  rotation data into a single entity. 
  Dual quaternions are created by quaternions if dual numbers 
  are used instead of real numbers as coefficients, i.e., they 
  are of the form $d:=A+\epsilon B$, where $A$ and $B$ 
  are ordinary quaternions and $\epsilon$ is the \emph{dual unit}, an 
  element that commutes with every element and satisfies 
  $\epsilon^2=0$ \cite{Kenwright:2012tl}. 
  A subset of these entities, called 
  \emph{unit dual quaternions}, are indeed isomorphic to the 
  transformation of a rigid body. A clear advantage of using 
  dual quaternions is the fact that we only need one framework 
  to maintain and that applying the encapsulated information 
  to a single point requires a simple sandwich operator. Moreover, 
  the rotation stored in the unit dual 
  quaternion $A+\epsilon B$ can be easily extracted as the  
  quaternion $r:=A$ is the unit quaternion 
  that amounts to the same rotation. Furthermore, if 
  $B^\star$ denotes the conjugate quaternion of $B$, 
  then $t:=2AB^\star$ is a pure quaternion whose coefficients 
  form the translation vector \cite{Kenwright:2012tl}.

\begin{figure}[t]
  \centering
  \includegraphics[width=0.95\textwidth]{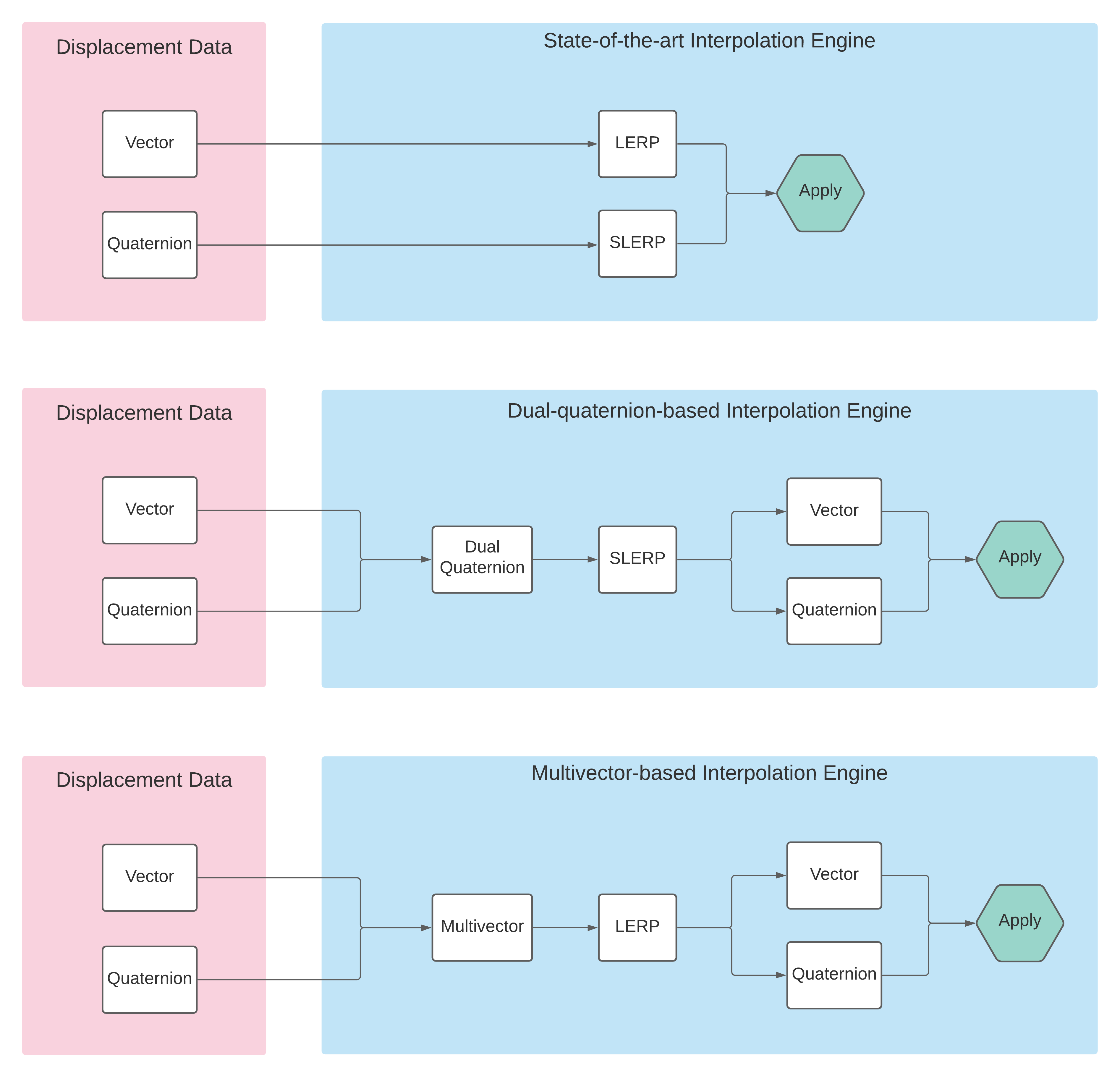}
  \caption{Algorithm layout of the different interpolation engines 
  used to generate intermediate frames.}
  \label{fig:sequence_diagrams}
  \end{figure}

  Taking advantage of the above, we propose the replacement 
  of the current state-of-the-art sequence 
  (see Figure~\ref{fig:sequence_diagrams},Top) with 
  the following (see 
  Figure~\ref{fig:sequence_diagrams},Middle). 
  The displacement data of an object is again represented as 
  a vector and a quaternion; in this way, only a total of 7 float 
  values (3 and 4 respectively) need to be transmitted. The VR engine 
  combines them in a dual quaternion \cite{Kenwright:2012tl}
  and interpolates with the previous state of the object, also 
  stored as a dual quaternion. Depending on the engine's and 
  the user's preferences, a number of in-between frames are 
  generated via SLERP interpolation \cite{Kavan2008} of the   
  the original and final data. For each dual-quaternion received 
  or generated, we decompose it to a vector and a quaternion and 
  apply them to the object. This step is necessary to take advantage
  of the built-in optimized mechanisms and GPU implementations 
  of the VR engine. 

  A major advantage of the proposed method is that we can obtain 
  similar results with the state-of-the-art method by sending less 
  keyframes per second. As an empirical law, we may send 20 
  displacement data per second with our method to obtain the same 
  quality of generated frames as if we had sent 30 data per second 
  with the current state-of-the-art method. This 33\% reduction 
  of required data applies for each user of the VR application, 
  greatly lowering the bandwidth required as more users join. 
  As an example, if $n$ users participate, the total displacement data 
  required for our method would be 1120$n$ bytes per second
  (assuming a float takes 8 bytes in a classic implementation) as
  opposed to 1680$n$ bytes per second with the default method. 
  The numbers of updates per second mentioned above relate 
  to the case of unrestricted-bandwidth network; for the 
  respective results regarding constrained networks Section~\ref{sec:our_results} and  Table~\ref{tbl:results}.
  The performance boost of our method is further validated as it 
  is used in the 
   MAGES SDK \cite{papagiannakis2020mages}
  for cooperative VR medical operations.

  The drawbacks of this method is the need to constantly transform 
  dual-quaternions to vector and rotation data after every 
  interpolation step but this performance overhead is tolerable
  as the extraction of the displacement data is accomplished 
  in a straight-forward way. 
  Also, performing SLERP on a dual quaternion 
  (proposed method) instead a quaternion (state-of-the-art method) 
  demands more operations per step. The trade-offs between 
  the two methods seem to favor our method, especially in the 
  case of collaborative VR applications.


  \subsection{Proposed Method Based on Multivectors} 
  \label{sub:proposed_method_based_on_multivectors}
  
\begin{figure}[t]
   \centering
   \includegraphics[width=0.95\textwidth]{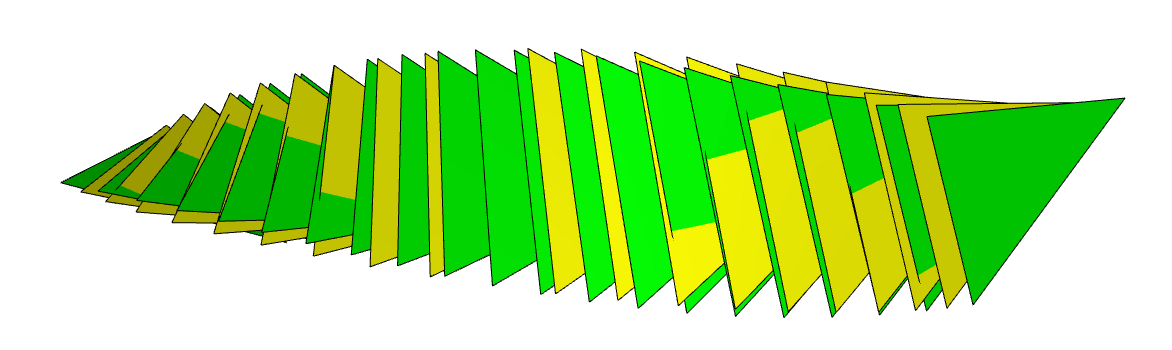}
   \caption{A triangular object is interpolated via multivectors. 
   A motor including both a translation and a rotation is applied 
   to the triangle via its mass center. Between the extreme positions
   of the object, 
   we generate 20 intermediate frames using LERP (yellow) and SLERP
   (green) interpolation of the multivector. Only minimal differences 
   are spotted between the two outcomes.}
   \label{fig:lerp_vs_slerp}
 \end{figure}

  The proposed method described in Section~\ref{sub:proposed_method_based_on_dual_quaternions} was based on the use 
  of dual quaternions and the fact that interpolating them (using SLERP)
  produced smooth intermediate frames. In this section, we go 
  one step further and suggest the use of multivectors instead of 
  dual-quaternions (see Figure~\ref{fig:sequence_diagrams},Bottom). 
  This transition can be done in a straight-forward way if we 
  use multivectors of 3D Conformal  
  (see \cite{DietmarFoundations}) or 3D Projective Algebra(see \cite{DorstBook} 
  and its updated Chapter 11 \cite{dorstguided} 
  found in \url{bivector.net}). The interpolation of the resulting 
  multivectors can be accomplished via LERP; if $M_1$ and $M_2$ 
  correspond to two consecutive displacement data, then we can 
  generate the in-between multivectors $(1-a)M_1+aM_2$, for as 
  many $a\in [0,1]$ as needed (and normalize them if needed). 
  Notice that since we are only 
  applying these displacements to rigid bodies, 
  we may use LERP instead of SLERP 
  (see Figure~\ref{fig:lerp_vs_slerp}); it is known that multivector 
  LERP does not yield correct results when applied to rigged models. 
  For every (normalized) multivector $M$ received or interpolated, 
  we may now 
  extract the translation vector and rotation quaternion. Assume that 
  $M=T*R$ where $T$ and $R$ are the multivectors encapsulating 
  the translation and rotation, we may extract them 
  depending on the Geometric Algebra used (all products below are 
  geometric unless stated otherwise):
  \begin{itemize}
    \item \textbf{3D PGA}: Given $M$, we evaluate $e_0M$.  
    Since in this algebra $T = 1 -0.5e_0(t_1e_1+t_2e_2+t_3e_3)$, 
    represents the translation by $(t_1,t_2,t_3)$ and $e_0e_0=0$, 
    it holds that $e_0M=e_0TR=e_0R$. Therefore, if 
    $e_0Q=e_0R=ae_0+be_{012}+ce_{013}+de_{023}$, we obtain 
    the multivector $R = a+be_{12}+ce_{13}+de_{23}$ which 
    corresponds to the quaternion $q=a-di+cj-bk$. 
    We can now evaluate $T$ as it equals 
    $MR^{-1}=M(a-be_{12}-ce_{13}-de_{23})=
    1 +xe_{01}+ye_{02}+ze_{03}$ and extract the 
    translation vector $(-2x,-2y,-2z)$. 
    \item \textbf{3D CGA}: Given $M$, we obtain $R$ by 
    adding the terms of $M$ that contain only the basis vectors  
     $\{1,e_1,e_2,e_3,e_{12},e_{23},e_{13}\}$. This derives 
     from the fact that $T=1-0.5*(t_1e_1+t_2e_2+t_3e_3)(e_4+e_5)$ 
     (which corresponds to the translation by $(t_1,t_2,t_3)$) 
     and therefore $TR=R+m$ where $m$ necessarily contains basis 
     elements containing $e_4$ and $e_5$ (or their geometric product) 
     that cannot be canceled out. After the obtaining of $R$, we 
     simply evaluate $T=MR^{-1}$, normalize it and extract
     the translation vector $(t_1,t_2,t_3)$ from the quantity 
     $t=T\cdot(e_5-e_4)=t_1e_1+t_2e_2+t_3e_3$. The conversion 
     of $R$ to quaternion and the evaluation of $R^{-1}$ is identical 
     with the case of 3D PGA above. 
  \end{itemize}

  The advantage of such a method lies on the fact that 
  we can use LERP blending of multivectors instead of SLERP. This 
  saves as a lot of time and CPU-strain; SLERP interpolation requires the 
  evaluation of a multivector's logarithm, which requires a lot 
  of complex operations \cite{Dorst:2011}. Notice that, LERP 
  is efficient in our case since 
  only rigid objects displacements are transfered via the network; 
  SLERP is the only efficient way for rigged models 
  animation via multivectors. Another gain of this proposed method is 
  the ability to incorporate it in an all-in-one GA framework, 
  that will use only multivectors to represent model, deformation and
  animation data. Such a framework is able to deliver 
  efficient results and embeds powerful modules 
  \cite{Papagiannakis:2013va,kamarianakis2021all,Papaefthymiou:2016dx}.

  The trade-offs of such an implementation are based on the fact 
  that modern VR engines do not natively support multivectors and 
  therefore production-ready modules, with basic functions implemented,
  are almost non-existent.
  An exception is the Klein C++ module for 3D PGA, found in 
  \url{www.jeremyong.com/klein}; for 3D CGA no such module is 
  available the moment this paper is written. This makes it 
  difficult for GA non-experts to adopt and implement such methods. 
  Furthermore, multivectors require 
  16 (3D PGA) or 32 (3D CGA) float values to be represented and 
  therefore even a simple addition between two amounts to 
  16 or 32 float operations respectively. 
  Unoptimized modules, usually running in CPU and not in GPU, 
  may result in slow rendering whereas. On the contrary, 
  optimized ones can take advantage 
  of the fact that very few of the multivector coordinates are 
  indeed non-zero, as the multivectors involved are always  
  motors, i.e., represent translations and/or rotations, and therefore
  have specific form.

\section{Our Results} 
\label{sec:our_results}

\begin{table}[t]
  \caption{Summary of the metrics of our methods (Ours) versus 
  the state-of-the-art methods (SoA). The first  column 
  describes the possible network quality which correlates 
  to the maximum number of updates per second that can be performed. 
  The second column contains the update rate required to 
  obtain the same QoE under the specific network quality limitations.
  The third column contains the comparison of the bandwidth  
  and the running time difference by our algorithms compared 
  with the SoA algorithm, when using the respective update 
  rates of the second column.}
  \begin{center}
  \begin{tabular}{|c|c|c|}
  \hline
  Network Quality & How to Achieve Best QoE & Metrics on Our Methods \\
  \hline
  \hline
  \multirow{2}{*}{Excellent} & SoA: 30 updates/sec & 33\% less bandwidth\\
  & Ours: 20 updates/sec & 16.5\% lower running time\\
  \hline 
  \multirow{2}{*}{Good} & SoA: 20 updates/sec & 50\% less bandwidth\\
  & Ours: 10 updates/sec & 16.5\% lower running time\\
  \hline 
  \multirow{2}{*}{Mediocre} & SoA: 15 updates/sec & 53\% less bandwidth\\
  & Ours: 7 updates/sec & 16.5\% lower running time\\
  \hline 
  \multirow{2}{*}{Poor} & SoA: 12 updates/sec & 58\% less bandwidth\\
  & Ours: 5 updates/sec & 16.5\% lower running time\\
  \hline 
  \end{tabular}
  \label{tbl:results}
  \end{center}
\end{table}

The methods proposed were implemented in Unity3D and applied to 
a VR collaborative training scenario. The video accompanying this
work demonstrates the effectiveness of our methods compared 
with the current state of the art. Specifically, we compare 
the three methods under different input rates per second, i.e., 
the keyframes send per second to the VR rendering engine. 
The input rates tested are 5,10,15 and 20 frames per second (fps),
where the last option is an optimal value 
to avoid CPU/GPU strain in collaborative VR scenarios. 
These rates are indicative values of the maximum possible fps 
that would be sent in a network whose bandwidth rates from 
very-limited (5 fps) to unrestricted (more than 20 fps). In lower fps, 
our methods yield jitter-less interpolated frames compared to 
the state-of-the-art method, which would require 30 fps to 
replicate similar output. As mentioned before, this 
reduction of required data that must be transfered per second 
by 33\%-58\% (depending on the network quality, see 
Table~\ref{tbl:results}) is multiplied by every active user, 
increasing the impact and the effectiveness of our methods in bandwidth-restricted environments.

\begin{figure}[t]
   \centering
   \includegraphics[width=0.95\textwidth]{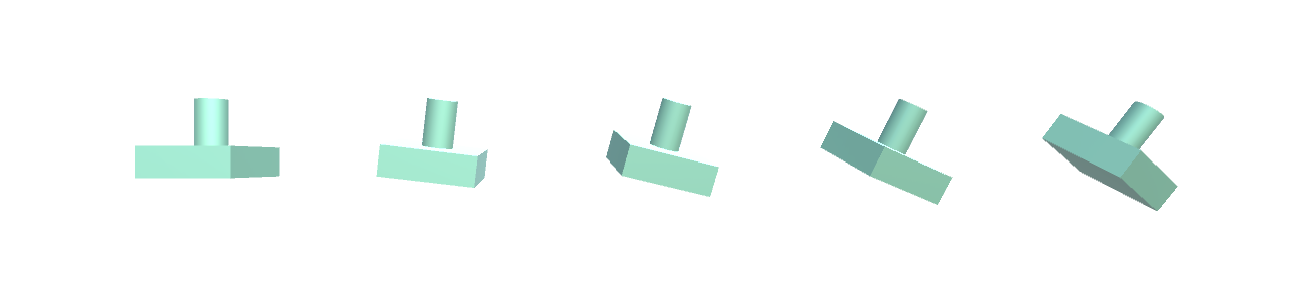}
   \includegraphics[width=0.95\textwidth]{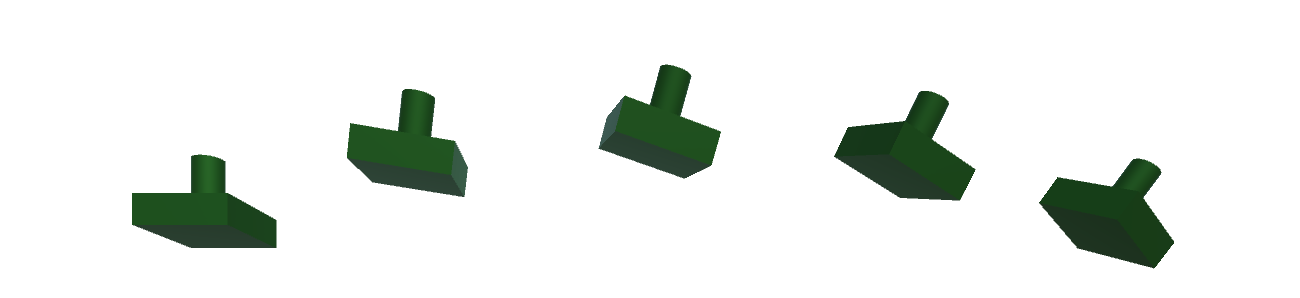}
   \includegraphics[width=0.95\textwidth]{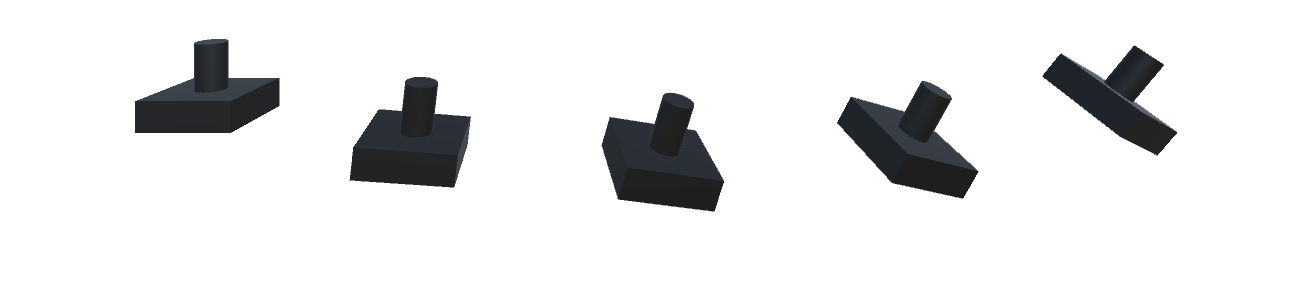}
   \caption{Different interpolation algorithms yield different, 
   yet jitter-less, intermediate frames. 
   (Top): State of the art: Vector and quaternion separate interpolation.
   (Middle): Dual-quaternion based interpolation algorithm.
   (Bottom): Multivector based interpolation algorithm.}
   \label{fig:interpolation}
 \end{figure}

The workflows of the two 
algorithms, compared with the current state of the art, are 
summarized in Figure~\ref{fig:sequence_diagrams}.
In Figure~\ref{fig:interpolation} we demonstrate the interpolation 
of the same object, at specific time intervals, for all methods; 
the intermediate frames feel natural for both methods proposed. 

In Table~\ref{tbl:results}, it is demonstrated that, 
under various network restrictions, both
proposed methods required less data (in terms of updates per sec) 
to be transmitted via the network to achieve the same QoE.
This decrease in data transfer leads to a lower energy 
consumption of the HMDs by 10\% (on average, preliminary result) 
and therefore enhances 
the overall mobility of the devices relying on batteries. 
Our methods provide a performance boost, decrease the required time
to perform the same deformation, with less keyframes but the same 
number of total generated frames, by 16.5\% 
(on average). The running times were produced in a PC with
a 3,1 GHz 16-Core Intel Core i9 processor, with 32 GBs of DDR4 memory. 
The same percentage of performance boost is expected in 
less powerful CPUs; in this case, the overall impact, in terms of 
absolute running time, will be even more significant.

\section{Conclusions and Future Work} 
\label{sec:conclusions}

In this work, we proposed two alternative interpolation algorithms 
based on dual-quaternions and multivectors respectively. These 
algorithms can be applied 
in the context of a networked virtual environment
to efficiently handle the interpolation of displacement data 
for hand-based VR HMDs.
The amount of displacement data per second that should be 
transmitted over the network to support a good QoE can 
be reduced using our methods instead of 
the state-of-the-art. This results in a performance boost and 
also lowers device energy consumption. The significance of our 
proposed methods are further highlighted in bandwidth-restricted
networks and when multiple users are involved. 
Our results are illustrated in a modern game engine and a medical VR
collaborative training scenario. 

The proposed algorithms and results can be further improved 
by using optimized C\# Geometric Algebra bindings 
(such as the ones provided in \url{bivector.net}). This would 
allow for efficient SLERP for the multivector interpolation 
engine and therefore unlock the potential to 
apply motors for rigged model animation in VR, as in 
\cite{Papagiannakis:2013va}. 
It is our intention
to integrate the algorithms proposed to an all-in-one GA framework 
that also enables features such as  
cut, tear and drill, as in 
\cite{kamarianakis2021all}.

\section{Acknowledgments  } 
\label{sec:Acknowledgments}

This work was co‐financed by European Regional Development 
Fund of the European Union and Greek national funds through the 
Operational Program Competitiveness, Entrepreneurship and Innovation, 
under the call RESEARCH – CREATE - INNOVATE 
(project codes:T1EDK-01149 and T1EDK-01448). 
The project also received funding from the European Union’s
Horizon 2020 research and innovation programme under grant agreement 
No 871793.


\bibliographystyle{splncs04}
\bibliography{references2}

\end{document}